\documentclass[%
  reprint,
  amsmath,amssymb,
  aps,
  ]{revtex4-1}

  \usepackage{graphicx}
  \usepackage{dcolumn}
  \usepackage{bm}
  \usepackage{multirow}
  \usepackage{adjustbox}
  \usepackage{tabularx}
  \newcolumntype{C}{>{\centering\arraybackslash}X}
  \newcolumntype{L}{>{\raggedright\arraybackslash}X}
  \newcolumntype{R}{>{\raggedleft\arraybackslash}X}

  \newcommand{\MKNN}{M_{\!\bar{K}\!N\!N}}
  \newcommand{\KN}{\bar{K}\!N}
  \newcommand{\KNN}{\bar{K}\!N\!N}
  \newcommand{\SNN}{\Sigma N\!N}
  \newcommand{\YNN}{Y\!N\!N}
  \newcommand{\YN}{Y\!N}
  
  \newcommand{\QF}{{\rm QF}_{\!\bar{K}\mathchar`-\!\,a\!\,b\!\,s}}
  \newcommand{\NA}{{\rm N\!A}}
  \newcommand{\m}{m_{X}}
  \newcommand{\q}{q_{X}}
  \newcommand{\BR}{B\!R}
  \newcommand{\valM}{2.328\pm0.003({\rm stat.})^{+0.004}_{-0.003}({\rm syst.})\ {\rm GeV}/c^{2}}
  \newcommand{\valB}{42\pm3({\rm stat.})^{+3}_{-4}({\rm syst.})\ {\rm MeV}}
  \newcommand{\valG}{100\pm7({\rm stat.})^{+19}_{-9}({\rm syst.})\ {\rm MeV}}
  \newcommand{\valQ}{383\pm11({\rm stat.})^{+4}_{-1}({\rm syst.})\ {\rm MeV}/c}
  \newcommand{\valCSLp}{9.3\pm0.8({\rm stat.})^{+1.4}_{-1.0}({\rm syst.})\ \mu{\rm b}}
  \newcommand{\valCSSp}{5.3\pm0.4({\rm stat.})^{+0.8}_{-0.6}({\rm syst.})\ \mu{\rm b}}
  \newcommand{\valBR}{1.7}

  \newcommand{\RIKEN}{$^1$}
  \newcommand{\RCNP}{$^2$}
  \newcommand{\Victoria}{$^3$}
  \newcommand{\Seoul}{$^4$}
  \newcommand{\IFINHH}{$^5$}
  \newcommand{\SMI}{$^6$}
  \newcommand{\Torino}{$^7$}
  \newcommand{\TorinoU}{$^8$}
  \newcommand{\Frascati}{$^{9}$}
  \newcommand{\Osaka}{$^{10}$}
  \newcommand{\Kyoto}{$^{11}$}
  \newcommand{\Tokyo}{$^{12}$}
  \newcommand{\OsakaE}{$^{13}$}
  \newcommand{\KEK}{$^{14}$}
  \newcommand{\TITEC}{$^{15}$}
  \newcommand{\TUM}{$^{16}$}
  \newcommand{\komaba}{$^{17}$}
  \newcommand{\Tohoku}{$^{18}$}
  \newcommand{\ECUTUM}{$^{19}$}
  \newcommand{\KIRAMS}{$^{20}$}
  \newcommand{\JAEA}{$^{21}$}
  \newcommand{\Sweden}{$^{22}$}
  \newcommand{\ELPH}{$^{23}$}
  \newcommand{\Chubu}{$^{24}$}
  \newcommand{\CREF}{$^{25}$}

  \usepackage[mathlines]{lineno}

\begin{document}

  \preprint{APS/123-QED}

  \title{Observation of a $\bm{\KNN}$ bound state in the $\bm{^{3}{\rm He} (K^{-}, \Lambda p)n}$ reaction}

  \author{
    T.~Yamaga\RIKEN
  }
  \email{takumi.yamaga@riken.jp}

  \author{
    S.~Ajimura\RCNP
  }
  \author{
    H.~Asano\RIKEN
  }
  \author{
    G.~Beer\Victoria
  }
  \author{
    H.~Bhang\Seoul
  }
  \author{
    M.~Bragadireanu\IFINHH
  }
  \author{
    P.~Buehler\SMI
  }
  \author{
    L.~Busso\Torino$^,$\TorinoU
  }
  \author{
    M.~Cargnelli\SMI
  }
  \author{
    S.~Choi\Seoul
  }
  \author{
    C.~Curceanu\Frascati
  }
  \author{
    S.~Enomoto\KEK
  }
  \author{
    H.~Fujioka\TITEC
  }
  \author{
    Y.~Fujiwara\Tokyo
  }
  \author{
    T.~Fukuda\OsakaE
  }
  \author{
    C.~Guaraldo\Frascati
  }
  \author{
    T.~Hashimoto\JAEA
  }
  \author{
    R.~S.~Hayano\Tokyo
  }
  \author{
    T.~Hiraiwa\RCNP
  }
  \author{
    M.~Iio\KEK
  }
  \author{
    M.~Iliescu\Frascati
  }
  \author{
    K.~Inoue\RCNP
  }
  \author{
    Y.~Ishiguro\Kyoto
  }
  \author{
    T.~Ishikawa\Tokyo
  }
  \author{
    S.~Ishimoto\KEK
  }
  \author{
    K.~Itahashi\RIKEN
  }
  \author{
    M.~Iwai\KEK
  }
  \author{
    M.~Iwasaki\RIKEN
  }
  \email{masa@riken.jp}
  \author{
    K.~Kanno\Tokyo
  }
  \author{
    K.~Kato\Kyoto
  }
  \author{
    Y.~Kato\RIKEN
  }
  \author{
    S.~Kawasaki\Osaka
  }
  \author{
    P.~Kienle\TUM
  }
  \thanks{deceased}
  \author{
    H.~Kou\TITEC
  }
  \author{
    Y.~Ma\RIKEN
  }
  \author{
    J.~Marton\SMI
  }
  \author{
    Y.~Matsuda\komaba
  }
  \author{
    Y.~Mizoi\OsakaE
  }
  \author{
    O.~Morra\Torino
  }
  \author{
    T.~Nagae\Kyoto
  }
  \author{
    H.~Noumi\RCNP$^,$\KEK
  }
  \author{
    H.~Ohnishi\ELPH
  }
  \author{
    S.~Okada\Chubu
  }
  \author{
    H.~Outa\RIKEN
  }
  \author{
    K.~Piscicchia\CREF$^,$\Frascati
  }
  \author{
    Y.~Sada\ELPH
  }
  \author{
    A.~Sakaguchi\Osaka
  }
  \author{
    F.~Sakuma\RIKEN
  }
  \author{
    M.~Sato\KEK
  }
  \author{
    A.~Scordo\Frascati
  }
  \author{
    M.~Sekimoto\KEK
  }
  \author{
    H.~Shi\SMI
  }
  \author{
    K.~Shirotori\RCNP
  }
  \author{
    D.~Sirghi\Frascati$^,$\IFINHH
  }
  \author{
    F.~Sirghi\Frascati$^,$\IFINHH
  }
  \author{
    S.~Suzuki\KEK
  }
  \author{
    T.~Suzuki\Tokyo
  }
  \author{
    K.~Tanida\JAEA
  }
  \author{
    H.~Tatsuno\Sweden
  }
  \author{
    M.~Tokuda\TITEC
  }
  \author{
    D.~Tomono\RCNP
  }
  \author{
    A.~Toyoda\KEK
  }
  \author{
    K.~Tsukada\Tohoku
  }
  \author{
    O.~Vazquez~Doce\Frascati$^,$\TUM
  }
  \author{
    E.~Widmann\SMI
  }
  \author{
    T.~Yamazaki\Tokyo$^,$\RIKEN
  }
  \author{
    H.~Yim\KIRAMS
  }
  \author{
    Q.~Zhang\RIKEN
  }
  \author{
    J.~Zmeskal\SMI
  }

  \collaboration{J-PARC E15 Collaboration}

  \affiliation{
    \RIKEN RIKEN Cluster for Pioneering Research, RIKEN, Wako, 351-0198, Japan 
  }
  \affiliation{
    \RCNP Research Center for Nuclear Physics (RCNP), Osaka University, Osaka, 567-0047, Japan
  }
  \affiliation{
    \Victoria Department of Physics and Astronomy, University of Victoria, Victoria BC V8W 3P6, Canada 
  }
  \affiliation{
    \Seoul Department of Physics, Seoul National University, Seoul, 151-742, South Korea 
  }
  \affiliation{
    \IFINHH National Institute of Physics and Nuclear Engineering - IFIN HH, Romania
  }
  \affiliation{
    \SMI Stefan-Meyer-Institut f\"{u}r subatomare Physik, A-1090 Vienna, Austria
  }
  \affiliation{
    \Torino Istituto Nazionale di Fisica Nucleare (INFN) Sezione di Torino, Torino, Italy
  }
  \affiliation{
    \TorinoU Dipartimento di Fisica Generale, Universita' di Torino, Torino, Italy
  }
  \affiliation{
    \Frascati Laboratori Nazionali di Frascati dell' INFN, I-00044 Frascati, Italy
  }
  \affiliation{
    \Osaka Department of Physics, Osaka University, Osaka, 560-0043, Japan
  }
  \affiliation{
    \Kyoto  Department of Physics, Kyoto University, Kyoto, 606-8502, Japan
  }
  \affiliation{
    \Tokyo  Department of Physics, The University of Tokyo, Tokyo, 113-0033, Japan
  }
  \affiliation{
    \OsakaE Laboratory of Physics, Osaka Electro-Communication University, Osaka, 572-8530, Japan
  }
  \affiliation{
    \KEK High Energy Accelerator Research Organization (KEK), Tsukuba, 305-0801, Japan
  }
  \affiliation{
    \TITEC Department of Physics, Tokyo Institute of Technology, Tokyo, 152-8551, Japan
  }
  \affiliation{
    \TUM Technische Universit\"{a}t M\"{u}nchen, D-85748, Garching, Germany
  }
  \affiliation{
    \komaba Graduate School of Arts and Sciences, The University of Tokyo, Tokyo, 153-8902, Japan
  }
  \affiliation{
    \Tohoku Department of Physics, Tohoku University, Sendai, 980-8578, Japan
  }
  \affiliation{
    \ECUTUM Excellence Cluster Universe, Technische Universit\"{a}t M\"{u}nchen, D-85748, Garching, Germany
  }
  \affiliation{
    \KIRAMS Korea Institute of Radiological and Medical Sciences (KIRAMS), Seoul, 139-706, South Korea
  }
  \affiliation{
    \JAEA ASRC, Japan Atomic Energy Agency, Ibaraki 319-1195, Japan 
  }
  \affiliation{
    \Sweden Department of Chemical Physics, Lund University, Lund, 221 00, Sweden
  }
  \affiliation{
    \ELPH Research Center for Electron Photon Science (ELPH), Tohoku University, Sendai, 982-0826, Japan
  }
  \affiliation{
    \Chubu Engineering Science Laboratory, Chubu University, Aichi, 487-8501, Japan
  }
  \affiliation{
    \CREF Centro Fermi-Museo Storico della Fisica e Centro studi e ricerche "Enrico Fermi", 000184 Rome, Italy
  }

  \date{\today}

  \begin{abstract}
    We have performed an exclusive measurement of the $K^{-}+\! ~^{3}{\rm He} \to \Lambda pn$ reaction at an incident kaon momentum of $1\ {\rm GeV}/c$.
    In the $\Lambda p$ invariant mass spectrum, a clear peak was observed below the mass threshold of $\bar{K}\!+\!N\!+\!N$, as a signal of the kaonic nuclear bound state, $\KNN$.
    The binding energy, decay width, and $S$-wave Gaussian reaction form-factor of this state were observed to be $B_{K} = \valB$, $\Gamma_{K} = \valG$, and $Q_{K} = \valQ$, respectively. 
    The total production cross-section of $\KNN$, determined by its $\Lambda p$ decay mode, was $\sigma^{tot}_{K} \cdot \BR_{\Lambda p} = \valCSLp$.
    We estimated the branching ratio of the $\KNN$ state to the $\Lambda p$ and $\Sigma^{0}p$ decay modes as $\BR_{\Lambda p}/\BR_{\Sigma^{0}p} \sim \valBR$, by assuming that the physical processes leading to the $\SNN$ final states are analogous to those of $\Lambda pn$.

  \end{abstract}

  \maketitle



  \section{\label{sec:introduction}
  Introduction
  }
  The bound system of an anti-kaon ($\bar{K}$) and a nucleon ($N$) has been studied ever since the $\Lambda(1405)$ was suggested as a $\bar{K}N$ molecular state \cite{dalitz.ap.1959,dalitz.pr.1967}.
  Based on numerous theoretical calculations of the chiral SU(3) dynamics and lattice-QCD, the interpretation that the $\Lambda(1405)$ has an internal structure as a $\KN$ molecular-state rather than a three-quark baryon has gained stronger theoretical support \cite{miyahara.prc.2016,kamiya.prc.2016,hall.prl.2015}.

  The possibility of a more general system containing a $\bar{K}$, called a kaonic nucleus, has also been discussed.
  Much theoretical work on these kaonic nuclei, especially in the $\KNN$ bound state, has been undertaken with various $\bar{K}N$ interaction models and calculation methods \cite{yamazaki.plb.2002,akaishi.prc.2002,ikeda.prc.2007,shevchenko.prl.2007,shevchenko.prc.2007,dote.npa.2008,wycech.prc.2009,dote.prc.2009,ikeda.prc.2009,barnea.plb.2012,bayar.prc.2013,revai.prc.2014,sekihara.ptep.2016,dote.prc.2017,ohnishi.prc.2017,dote.plb.2018}.
  The $\KNN$ bound state has charge $+1$ and isospin $I=1/2$ (symbolically denoted as $K^{-}pp$ for the $I_{z}=+1/2$ state) and its spin and parity are considered to be $J^{P}=0^{-}$.
  The existence of the $\KNN$ bound state is generally supported by all the calculations mentioned above; however, the estimated binding energies and widths of the state are widely spread.

  To search for the $\KNN$ bound state, we conducted the experiment J-PARC E15 using the in-flight $K^{-}$ beam at J-PARC.
  In the first measurement of the experiment, we demonstrated a significant yield excess well below the $\KNN$ mass threshold ($\MKNN = m_{\bar{K}}+2m_{N}\sim2.37\ {\rm GeV}/c^{2}$) in the inclusive analysis of the $^{3}{\rm He}(K^{-}, n)$ reaction \cite{hashimoto.ptep.2015}, which suggests the strongly attractive nature of the $\bar{K}N$ interaction.
  We therefore extended the analysis focusing on the simplest exclusive channel, the $\Lambda pn$ final state, which consists of three baryons including the lightest hyperon \cite{sada.ptep.2016}.
  Because $s$-quark conservation is secured in nuclear reactions governed by the strong interaction, we can trace the $s$-quark flow.
  Thus, the interaction between a recoiled $\bar{K}$ and two spectator nucleons, $\bar{K}$--$\!N\!N$, can be studied by $\YN$-pair analysis, which will tell us the reaction dynamics and formation signature of $\KNN$, if it exists.
  As described in Ref.~\cite{sada.ptep.2016}, a kinematical anomaly, a concentration of events around $\MKNN$, was observed only in the $\Lambda p$ invariant mass spectrum. 
  To study this anomaly, we performed a second measurement and found a peak structure in the $\Lambda p$ invariant mass spectrum located below $\MKNN$, which we interpreted as a signal of the $\KNN$ bound state \cite{ajimura.plb.2019}.

  In Ref.~\cite{ajimura.plb.2019}, the $\Lambda pn$ final state was selected by detecting $\Lambda p$ and by the kinematical consistency of the reaction including a missing neutron.
  However, we cannot entirely exclude the two final states $\Sigma^{0}pn$ and $\Sigma^{-}pp$ by the selection. 
  We treated the effect of the contamination of the $\SNN$ final state (the $\Sigma N$ decay channel of $\KNN$) as a source of systematic error for simplicity.
  In this article, we evaluated the effect of the $\SNN$ final state contamination and estimated the $\KNN$ decay branch to the $\Sigma^{0}p$ channel in a self-consistent way.


  \section{\label{sec:j-parc-e15-experiment}
  J-PARC E15 experiment
  }
  We measured the $K^{-} +\! ~^{3}{\rm He} \to \Lambda pn$ reaction to search for the $\KNN$ bound state by its $\Lambda p$ decay mode.
  The incident momentum of the $K^{-}$ beam is chosen to be $p_{K} = 1\ {\rm GeV}/c$ to maximize the cross-section of the elementary $K^{-}N \to \bar{K}N$ reaction, corresponding to $\sqrt{s}=1.8\ {\rm GeV}$.

  Because the kinematical anomaly was found only in the $\Lambda p$ invariant mass of the $\Lambda pn$ final state, we analyzed the process as two successive reactions, {\it i.e.}, 
  \begin{equation}
    \label{eq:reaction}
    \begin{split}
      K^{-}+~^{3}{\rm He} \to &X + n,\\
      &X \to \Lambda p.
    \end{split}
  \end{equation}
  The former two-body reaction can be characterized by two parameters, the invariant mass of $X$ ($m_{X}$) and momentum transfer to $X$ ($q_{X}$).
  We interpret the $X$ formation reaction in a more microscopic way, described in the framework of the cascade reactions
  \begin{equation}
    \label{eq:X-formation}
    \begin{split}
      K^{-}+N \to &\bar{K} + n,\\
      &\bar{K}+NN \to X,
    \end{split}
  \end{equation}
  in which a virtual kaon $\bar{K}$ is produced in the primary reaction between a $K^{-}$ and a nucleon followed by a formation reaction of the $X$ resonance together with two spectator nucleons.
  In the microscopic view, $m_{X}$ corresponds to the invariant mass of the $\bar{K}+N\!N$ system, and $q_{X}$ is the 3-momentum of the intermediate virtual $\bar{K}$ that can be measured by the momentum of $\Lambda + p$ in the final state in the laboratory frame.
  At $p_{K} = 1\ {\rm GeV}/c$, the minimum $q_{X}$ is as small as $\sim 200\ {\rm MeV}/c$ when the neutron is formed in the forward direction, so we can expect a large $\bar{K}$ sticking probability to the two residual nucleons.

  The experiment was performed at the hadron experimental facility of J-PARC.
  A high-intensity secondary $K^{-}$ beam, produced by bombarding a primary gold target with a 30-GeV proton beam, is transported along the K1.8BR beam line.
  Other secondary particles in the beam are removed by an electrostatic separator.

  A beam-line detector system and a cylindrical detector system (CDS) are used to measure incident $K^{-}$ and scattered charged particles, respectively.
  A detailed description of the experimental setup is given in Refs.~\cite{agari.ptep.2012.02B009,agari.ptep.2012.02B011,iio.nim.2012}; however, we summarize the basics as follows.

  The beam-line detector system measures the time of flight and momentum of the $K^{-}$ beam.
  At the on-line level, $K^{-}$ is identified by an aerogel Cherenkov detector.
  The position and direction of the beam are measured by a drift chamber located just in front of the experimental target of liquid $^{3}{\rm He}$.
  The liquid $^{3}{\rm He}$ target is located at the final focus point of the beam line.
  The target cell of $^{3}{\rm He}$ has a cylindrical shape with a diameter of 68 mm and a length along the beam direction of 137 mm, and has a density of $\sim 80\ {\rm mg}/{\rm cm^{3}}$.
  We accumulated $^{3}{\rm He}$-filled data as the experimental run, and the empty target data as a background study.
  The CDS surrounding the $^{3}{\rm He}$ target is composed of a cylindrical drift chamber and a cylindrical hodoscope.
  The detectors are installed inside a solenoid magnet to measure the momenta of the scattered charged particles.

  \section{\label{sec:analysis}
  Analysis
  }
  Particle identification and momentum reconstruction of the $K^{-}$ beam and scattered charged-particles were performed.
  Then, the $K^{-}+~^{3}{\rm He} \to \Lambda pn$ final state was selected, where $\Lambda$ and $p$ were detected by CDS and the missing-$n$ was identified kinematically.
  For the selected $\Lambda pn$ events, we measured a 2D distribution of the invariant mass of the $\Lambda p$ and the momentum transfer to the $\Lambda p$.
  To investigate the production of the $\KNN$ bound state, we conducted a spectral fitting to the 2D distribution.


  \subsection{\label{sec:analysis:beam-and-scattered}
  Beam and scattered particle analysis
  }
  For the $K^{-}$ beam, we applied time-of-flight-based PID selection to achieve a high purity of kaon identification.
  Contamination from the in-flight kaon decay was eliminated by checking the track inconsistency as a particle trajectory recorded by drift chambers.
  The beam momentum was determined with a second-order transfer matrix of the final beam-line dipole spectrometer magnet calculated using the TRANSPORT code \cite{transport}.
  A typical momentum resolution was estimated to be 0.2\%.

  The trajectories of the charged particles from the $K^{-}+~^{3}{\rm He}$ reaction were measured by the CDS.
  We designed the magnet to have sufficient magnetic uniformity in the effective region of the CDS to apply a simple helical fit to each trajectory to analyze its momentum.
  The absolute magnetic field strength was 0.715 T, calibrated using monochromatic invariant-mass peaks of $K^{0}_{s} \to \pi^{+}\pi^{-}$ and $\Lambda \to p\pi^{-}$ decays.
  The PID was conducted by a conventional method based on the 2D event distribution over the mass-square and momentum.
  In the present analysis, a $\pm \ 2.5\ \sigma$ region from the intrinsic mass was selected for each particle.
  Any overlap of two different PID regions was rejected to reduce miss-identification \cite{sada.ptep.2016}.
  The inefficiency due to the overlap rejection was corrected in the analysis efficiency.
  After the particle identification, an energy-loss correction was applied by considering all the materials on the trajectory of the particle to obtain its initial momentum.


  \subsection{\label{sec:analysis:event-selection}
  Event selection of $\bm{\Lambda pn}$ final state
  }
  To select the $K^{-}+~^{3}{\rm He} \to \Lambda pn$ reaction, three charged particles, $pp\pi^{-}$, were required.
  From the $pp\pi^{-}$, we examined two possible $p\pi^{-}$-pairs as for $\Lambda$ candidates ($\Lambda'$).
  A candidate trajectory is tentatively defined by the $p\pi^{-}$ vertex (the nearest point of the two trajectories) and synthetic momentum vector of the two.
  Then, we checked if the event kinematics is consistent with the $\Lambda pn$ final state, by a kinematical fitting. 
  In the kinematical fitting, the $p\pi^{-}$-pair invariant mass ($m_{p\pi^{-}}$) and the $pp\pi^{-}$ missing mass ($m_{R^{0}}$ in the $^{3}{\rm He}(K^{-}, pp\pi^{-})R^{0}$ reaction) are used to derive the $\chi^{2}$ (degrees of freedom = 2, in the present case) as an indicator of the kinematical consistency to be the $\Lambda pn$ final state.
  The ``{\it KinFitter}'' package based on the \textsc{Root} classes \cite{kinfitter} was used to search for the minimum $\chi^{2}$.

  To include geometrical consistency of the event topology in the consistency test, a log-likelihood $l(\bm{x})$ is introduced as
  \begin{equation}
    \label{eq:likelihood}
    l({\bm x}) = -{\rm ln} \prod_{i=1}^{5} p_{i}(x_{i}),
  \end{equation}
 where $p_{i}$ is the probability density function of the $i$-th variable estimated by a Monte Carlo simulation, and the maximum value is renormalized to be one, so as to make $l(\bm{x}) = 0$ at the most probable density point of the parameter set.
  $\bm{x}$ stands for
  \begin{equation}
    \label{eq:quantities}
    \bm{x} = \left( \chi^{2}, D_{K^{-}p}, D_{K^{-} \Lambda'}, D_{\Lambda' p}, D_{p\pi^{-}} \right),
  \end{equation}
  where the five variables are the $\chi^{2}$ given by the kinematical fitting, the distances of closest approach for incoming $K^{-}$ with $p$ ($D_{K^{-}p}$) and with $\Lambda'$ ($D_{K^{-} \Lambda'}$), the distance of closest approach of $\Lambda'$ and $p$ ($D_{\Lambda' p}$), and the minimum approach of the $p\pi^{-}$-pair at the $\Lambda'$ decay point ($D_{p\pi^{-}}$).
  Finally, both the $K^{-}\Lambda'$ and $K^{-}p$ vertices were required to be in the fiducial volume of the target, to reduce the background from the target cell.
  In this examination, more than 99.5\% of the $p\pi^{-}$ were paired correctly in the simulation.
%
  \begin{figure}[b]
    \includegraphics[width=1.0\linewidth]{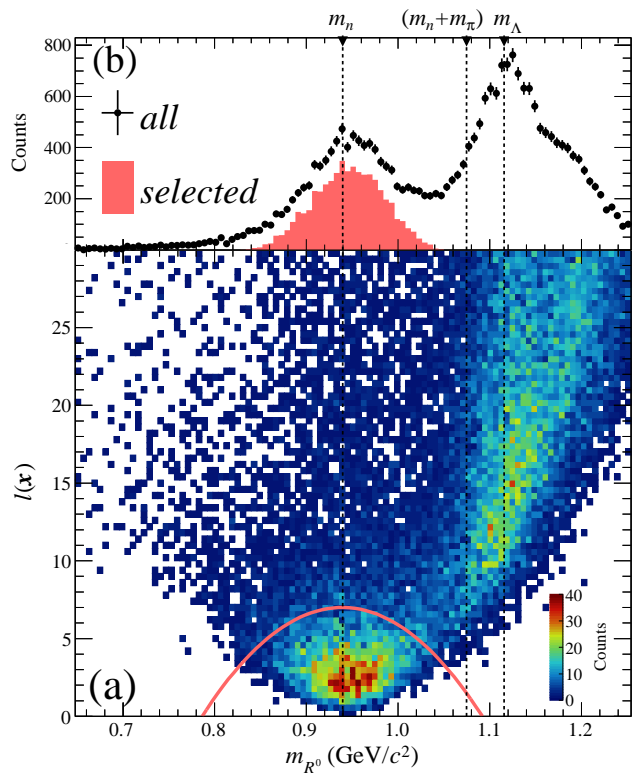}
    \caption{
      (color online) (a) 2D plot of $m_{R^{0}}$ in the $^{3}{\rm He}(K^{-},pp\pi^{-})R^{0}$ reaction, and $l(\bm{x})$.
      (b) Projected spectrum on the $m_{R^{0}}$ axis by selecting $l(\bm{x})<30$.
      The vertical black dashed lines are the masses of $n$ ($m_{n}$),  $N+\pi^{0}$ ($m_{N}+m_{\pi}$), and $\Lambda$ ($m_{\Lambda}$).
      Events from the $\Lambda pn$ final state make a strong event concentration at the bottom of the 2D plot, where $m_{R^{0}} \sim m_{n}$.
      The $\Lambda pn$ event was selected below the red line in the 2D plot.
      The projection of selected events is shown by the red histogram in (b).
    }
    \label{fig:mm_vs_lnl}
  \end{figure}
  \begin{figure}[t]
    \includegraphics[width=1.0\linewidth]{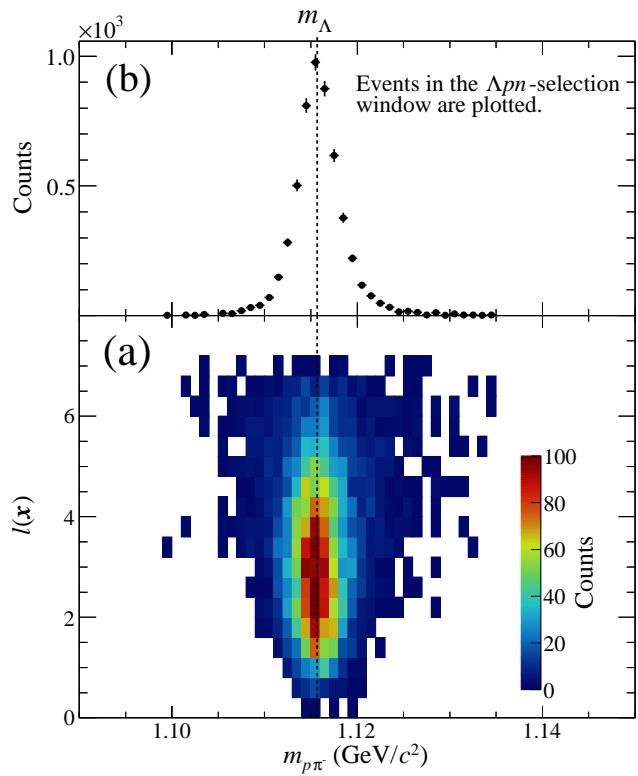}
    \caption{
      (color online) (a) 2D plot of $m_{p\pi^{-}}$ and $l(\bm{x})$.
      (b) Projected spectrum on the $m_{p\pi^{-}}$ axis.
      Events in the $\Lambda pn$-selection window (shown in Fig.~\ref{fig:mm_vs_lnl}-(a)) are plotted.
      The vertical black dashed line is the $\Lambda$ mass.
    }
    \label{fig:impip_vs_lnl}
  \end{figure}
  \begin{figure*}[t]
    \includegraphics[width=1.0\linewidth]{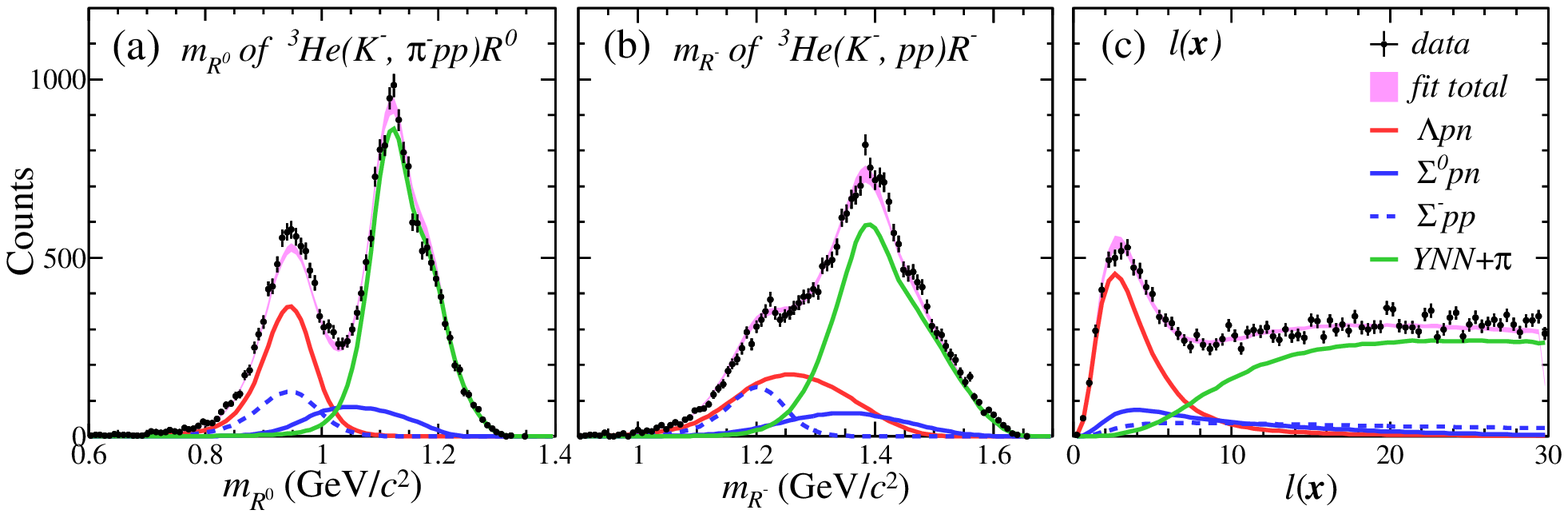}
    \caption{
      (color online) Distributions of (a) $m_{R^{0}}$ of $^{3}{\rm He}(K^{-}, \pi^{-}pp)R^{0}$ (the same as Fig.~\ref{fig:mm_vs_lnl}-(b)), (b) $m_{R^{-}}$ of $^{3}{\rm He}(K^{-}, pp)R^{-}$, and (c)  $l(\bm{x})$.
      For the $m_{R^{0}}$ and $m_{R^{-}}$ spectra, $l(\bm{x})<30$ was selected.
      These three distributions were simultaneously fitted by simulated spectra shown by colored lines.
      The fitting chi-square and number of degrees of freedom were 917 and 506, respectively.
      For mesonic ($\YNN+\pi$), the final states all of possible charged states and combinations were summed.
    }
    \label{fig:mmlnl}
  \end{figure*}

  The event distribution of $m_{R^{0}}$ and $l(\bm x)$ is shown as a 2D plot in Fig.~\ref{fig:mm_vs_lnl}-(a).
  A strong event concentration is seen at the bottom of the figure, which corresponds to the non-mesonic $\Lambda pn$ final state. 
  As shown in the $m_{R^{0}}$ spectrum, Fig.~\ref{fig:mm_vs_lnl}-(b), $\Lambda pn$ events make a clear peak at $m_{n}$, and the events are clearly separated from the mesonic ($\YNN+\pi$) final states located at $m_{R^{0}} > m_{N}+m_{\pi}$.
  To improve the $\Lambda pn$-selection, we selected $\Lambda pn$ events on the 2D plane of $m_{R^{0}}$ and $l(\bm{x})$, as indicated by the red line in Fig.~\ref{fig:mm_vs_lnl}-(a).

  The 2D plot of $m_{p\pi^{-}}$ and $l(\bm{x})$, applying the $\Lambda pn$-selection window, is shown in Fig.~\ref{fig:impip_vs_lnl}-(a), and the projection onto $m_{p\pi^{-}}$ is shown in Fig.~\ref{fig:impip_vs_lnl}-(b).
  As shown in the figure, $\Lambda$ is clearly selected.
  The tail of the $\Lambda$-peak is quite small; however, we should note that it does not secure the purity of the $\Lambda pn$ final state, in that the tail is removed by the kinematical fitting procedure through the $\chi^{2}$ evaluation.
  In the present $\Lambda pn$-selection, the other final states may come in, as is indicated in Fig.~\ref{fig:mm_vs_lnl}-(a).

  To evaluate the contamination yields of the other final states, we conducted a detailed simulation as shown in Fig.~\ref{fig:mmlnl}.
  In this simulation, we generated non-mesonic $\YNN$ final states ($\Lambda pn$, $\Sigma^{0}pn$, and $\Sigma^{-}pp$) according to the fit result (described in Sec.~\ref{sec:results-and-discussion:2d-fitted-spectra}) to make the simulation realistic.
  For simplicity, the event distribution of mesonic final states, which make smaller contributions to the $\Lambda pn$-selection window, are generated proportional to the phase space.

  As shown in Fig.~\ref{fig:mmlnl}-(a), it is difficult to eliminate the $\Sigma^{0} pn$ and $\Sigma^{-}pp$ final state events in the $\Lambda pn$-selection window, since the $m_{R^{0}}$ spectra of contaminations of the two components are very similar.
  In particular, the $\Lambda pn$ and $\Sigma^{-}pp$ final states have the same $m_{R^{0}}$ distribution.
  This is because $R^{0} = n$, $\gamma + n$, and $n$ for the $\Lambda pn$, $\Sigma^{0}pn$, and $\Sigma^{-}pp$ final states, respectively.
  Thus, we plotted the $m_{R^{-}}$ spectrum of $^3{\rm He}(K^{-}, pp)R^{-}$, as shown in Fig.~\ref{fig:mmlnl}-(b), to give $R^{-} = \pi^{-} + n$, $\pi^{-} + \gamma + n$, and $\Sigma^{-}$ for the $\Lambda pn$, $\Sigma^{0}pn$, and $\Sigma^{-}pp$ final states, respectively.
  As shown in the figure, the relative yields can be evaluated easily, since the $\Sigma^{-}pp$ final state makes a peak at the $\Sigma^{-}$ intrinsic mass, while the $\Lambda pn$ final state becomes even broader in the $m_{R^{-}}$ distribution.
  Figure~\ref{fig:mmlnl}-(c) is the projection of the events onto $l(\bm{x})$, where the $\Lambda pn$ final state has smaller $l(\bm{x})$ than the other final states.

  The relative yields of the signal and contaminations in the present $\Lambda pn$-selection window were estimated by the simultaneous fitting of these three spectra.
  The result is summarized in Tab.~\ref{tab:relative-yield}.
  The fit result improved substantially by applying realistic $\Lambda pn$ distribution, together with $\Sigma^{0}pn$ and $\Sigma^{-}pp$ contributions to the spectra.
  However, the fit result, chi-square 917 over degrees of freedom 506 of Fig.~\ref{fig:mmlnl}, might not be very sufficient by number.
  This is because we accepted events having relatively large $l(\bm{x})$ to evaluate the contamination from the mesonic final states, whose distribution is simply assumed to be proportional to the phase space.
  Thus, the systematic uncertainties of the table were evaluated by limiting the fitting data region of Fig.~\ref{fig:mmlnl} to $l(\bm{x})<10$ to reduce the contamination effect from mesonic final states.

  Contaminations from the mesonic final state and from the $K^{-}$ reaction at the target cell are negligible. 
  Thus, we focused on the non-mesonic $\Sigma^{0}pn$ and $\Sigma^{-}pp$ final states ($\SNN$) in the following analysis (Sec.~\ref{sec:analysis:effect-of-SNN-contamination}).

  \begin{table}[h]
    \caption{
      Relative yields of signal and contaminations in the present $\Lambda pn$-selection.
      The first and second errors are statistical and systematic, respectively.
    }
    \label{tab:relative-yield}
    \begin{tabularx}{\linewidth}{LR} \hline \hline
      Source                              & Relative yield ($R_{j}$) (\%) \\
      \hline
      $\Lambda pn$ (signal)               & $76.3\pm1.6\pm0.5$ \\
      $\Sigma^{0} pn$                     & $12.0\pm0.8\pm0.6$ \\
      $\Sigma^{-} pp$                     & $ 7.1\pm0.3\pm1.4$ \\
      Total mesonic final states          & $ 1.5\pm0.1\pm0.4$ \\
      $K^{-}$ reaction at the target cell & $ 3.1\pm0.0\pm0.4$ \\
      \hline
      \hline
    \end{tabularx}
  \end{table}

  \subsection{\label{sec:analysis:m-and-q}
  $\bm{\m}$ and $\bm{\q}$ distributions	
  }
  For the $\Lambda pn$-selected events, we measured the invariant mass of the $\Lambda p$ system ($\m$) and the momentum transfer to the $\Lambda p$ system ($\q$).
  As shown in Eq.~\ref{eq:reaction}, $\q$ can be given by the momenta of $\Lambda$ ($\bm{p}_{\Lambda}$) and $p$ ($\bm{p}_{p}$) as
  \begin{equation}
    \q = | \bm{p}_{\Lambda} + \bm{p}_{p} |.
  \end{equation}
  \begin{figure}[t]
    \includegraphics[width=1.0\linewidth]{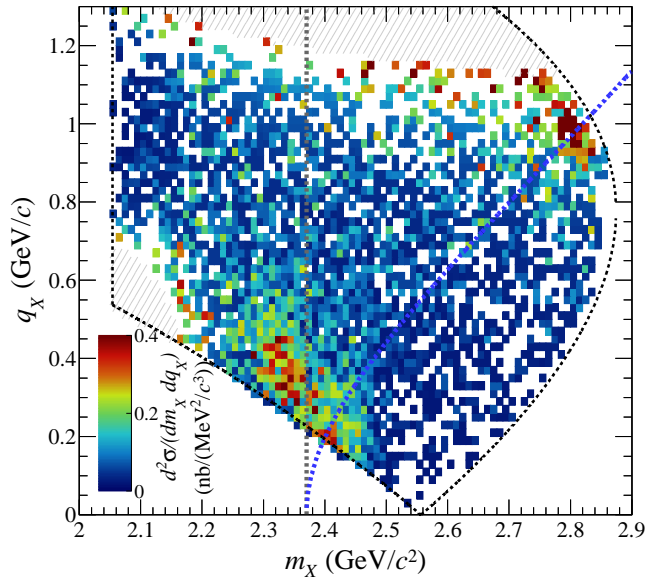}
    \caption{
      (color online) 2D plot on the $\m$ and $\q$ plane after acceptance correction.
      The black dotted line shows the kinematical limit of the reaction.
      The vertical gray dotted line and blue dotted curve are $\MKNN$ and $M_{F}(q)$, respectively.
      The gray hatched regions indicate where the experimental efficiency is $<0.5\%$.
    }
    \label{fig:im_vs_q}
  \end{figure}

\begin{figure*}[t]
    \includegraphics[width=0.7\linewidth]{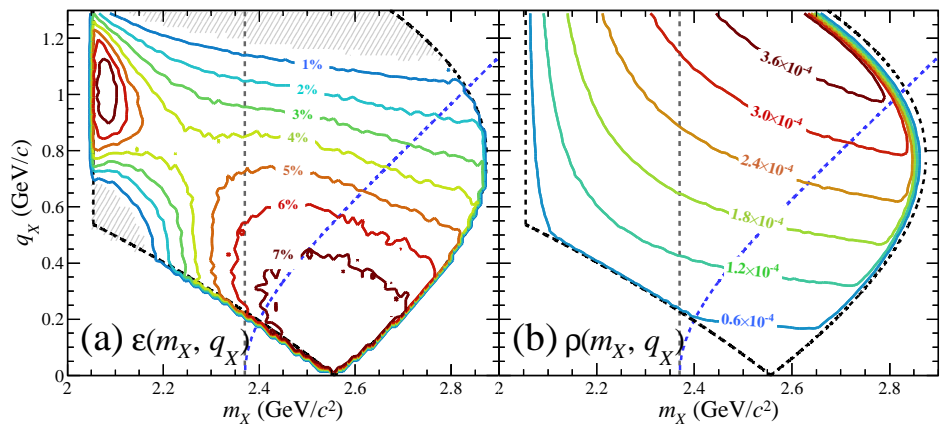}
    \caption{
      (color online)
      (a) Simulated spectra of experimental efficiency $\varepsilon(\m,\q)$ for $\Lambda pn$ final states.
      $\varepsilon(\m,\q)$ includes geometrical acceptance of CDS and analysis efficiency (decay branching ratio of $\Lambda$ is also taken into account). 
      The efficiency is calculated bin by bin.
      The hatched regions are insensitive in the present setup, where $\varepsilon < 0.5\%$.
      (b) Lorentz-invariant $\Lambda pn$ phase space $\rho(\m,\q)$ taking into account the kaon beam momentum bite.
      The ratio is normalized by one generated event.
      The roughness of the contours in both (a) and (b) is due to the limited statistics of the simulation. 
      The vertical gray dotted lines and blue dotted curves are the same as in Fig.~\ref{fig:im_vs_q}.
    }
    \label{fig:im_vs_q_effic}
  \end{figure*}

  Figure~\ref{fig:im_vs_q} shows the 2D event distribution on the $\m$ and $\q$ plane.
  As shown in the figure, there are very strong event-concentrating regions.
  To show these event-concentrations unbiased manner, an acceptance correction was applied to the data, to make the results independent of both the experimental setup and analysis code.
  The events density, represented by a color code, is given in units of the double differential cross-section:
  \begin{equation}
    \label{eq:differential-cross-section}
    \frac{d^{2}\sigma}{d\m\,d\q} = \frac{N(\m,\q)}{\varepsilon(\m,\q)} \frac{1}{\Delta \m} \frac{1}{\Delta \q} \frac{1}{\mathcal{L}},
  \end{equation}
  where $N(\m,\q)$ is the obtained event number in $\Delta \m=10\ {\rm MeV}/c^{2}$ and $\Delta \q=20\ {\rm MeV}/c$ (bin widths of $\m$ and $\q$, respectively).
  $\mathcal{L}$ is the integrated luminosity, evaluated to be $2.89 \pm 0.01\ {\rm nb}^{-1}$.
  $\varepsilon(\m,\q)$ is the experimental efficiency, which is quite smooth, as shown in Fig.~\ref{fig:im_vs_q_effic}-(a), around all the events-concentrating regions of Fig.~\ref{fig:im_vs_q}.

  After the acceptance correction, if no intermediate state, such as $X$, exists in the $K^{-} + ~^{3}{\rm He} \to \Lambda pn$ reaction, then the event distribution will simply follow the $\Lambda pn$ phase space $\rho(\m,\q)$ without having a specific form-factor as given in Fig.~\ref{fig:im_vs_q_effic}-(b).
  In contrast to the data in Fig.~\ref{fig:im_vs_q}, $\rho(\m,\q)$ is smooth for the entire kinematically allowed region.

  To account for the observed event distribution, three physical processes were introduced as in Ref.~\cite{ajimura.plb.2019}.
  Details of the physical processes, the formulation of each fitting function, and the fitting procedures are described in the following sections.

  \begin{figure*}[t]
    \includegraphics[width=1.0\linewidth]{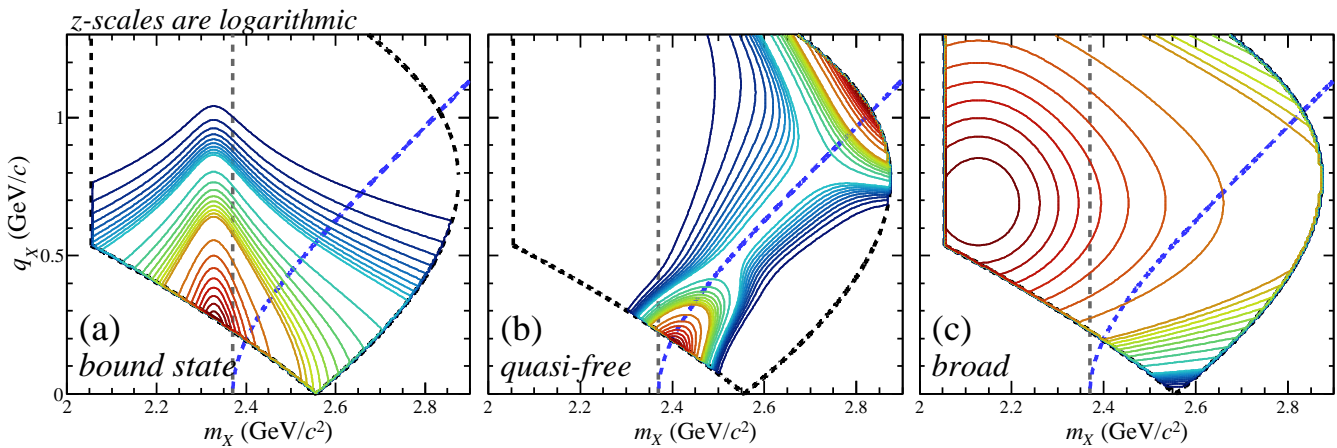}
    \caption{
      (color online) 2D spectral functions for (a) the $\KNN$ bound state $f_{K}(\m,\q)$ (see Eq.~\ref{eq:bound-state}), (b) the $\QF$ process $f_{F}(\m,\q)$ (see Eq.~\ref{eq:quasi-free}), and (c) a broad distribution $f_{B}(\m,\q)$ (see Eq.~\ref{eq:broad}). 
      For all the figures, the function strength is given in a logarithmic scale, where the contours are in the steps of 10\% (red--orange), 1\% (orange--cyan), and 0.1\% (cyan--blue) compared to the maximum density of each function.
      The vertical gray dotted lines and blue dotted curves are the same as in Fig.~\ref{fig:im_vs_q}.
    }
    \label{fig:fit_functions}
  \end{figure*}
  \subsection{\label{sec:analysis:spectral-fitting}
  2D model fitting functions
  }
  We considered the following three processes: $K$) the $\KNN$ bound state, $F$) the non-mesonic quasi-free kaon absorption ($\QF$) process, and $B$) a broad distribution covering the whole kinematically allowed region of the $\Lambda pn$ final state.
  To decompose those processes, we conducted 2D fitting for the event distribution. 

  The production yields of these three processes ($F_{i}(\m,\q)$ for $i=K,F,B$) observed in the $\Lambda pn$ final state should be proportional to the $\Lambda pn$ phase space $\rho(\m,\q)$.
  Thus, $F_{i}(\m,\q)$ can be described as the product of $\rho(\m,\q)$ and specific spectral terms for the $i$-th process of a component $f_{i}(\m,\q)$, as
  \begin{equation}
    F_{i}(\m,\q) = \rho(\m,\q) f_{i}(\m,\q).
  \end{equation}
  Figure~\ref{fig:fit_functions} shows typical 2D distributions of $f_{i}(\m,\q)$ for the three processes.
  All the parameters of the fitting functions described below are fixed to the final fitting values.

  To make $f_{i}$ automatically fulfill time-reversal symmetry, we limited ourselves to using $q_{X}$-even terms to formulate the fitting functions described below,
with one exception.
  The details and the reason for the exception are described below.

  \subsubsection{\label{sec:analysis:spectral-fitting:kbarnn}
  $\KNN$ production {\rm (}$i=K${\rm )}
  }
  As described in Ref.~\cite{ajimura.plb.2019}, we formulated the formation cross-section of the $\KNN$ bound state according to the reaction in Eq.~\ref{eq:reaction} with a plane-wave impulse approximation (PWIA) with a harmonic oscillator wave function.
  In this way, we simplified the microscopic reaction mechanism in Eq.~\ref{eq:X-formation}.
  We assumed that the spatial size of the bound state is much smaller than that of $^{3}{\rm He}$, so the size term of $^{3}{\rm He}$ was ignored in the formula.
  The time integral gives a Breit--Wigner formula in the $\m$-direction, and the spatial-integral gives a Gaussian-form factor as
  \begin{equation}
    \begin{split}
      \label{eq:bound-state}
      f_{K} \left( \m, \q \right) &=\frac{(\Gamma_{K}/2)^{2}}{ \left( \m - M_{K} \right) ^{2} + \left( \Gamma_{K}/2 \right)^{2} } \\
      &\quad \times  A_{0}^{K} \,\exp\! \left( - \frac{ \q^{2}}{Q_{K}^{2}} \right),
    \end{split}
  \end{equation}
  where $M_{K}$, $\Gamma_{K}$, and $Q_{K}$ are the mass, decay width, and $S$-wave reaction form-factor (involving microscopic reaction dynamics) parameter of the bound state, respectively.

  \subsubsection{\label{sec:analysis:spectral-fitting:quasi-free}
  Non-mesonic $\QF$ process {\rm (}$i=F${\rm )}
  }
  When the invariant mass $\m$ of the secondary reaction in Eq.~\ref{eq:X-formation} is larger than the threshold $\MKNN$, the recoil-kaon can behave as an approximately free particle; {\it i.e.}, $X$ can be any channel, such as $\bar{K}+N+N$, $Y+N$, or other mesonic channels.
  Among these, we denote the $Y+N$ channel as the non-mesonic $\QF$ process. 
  Specifically, $Y$ and $N$ are $\Lambda$ and $p$ in the $\Lambda pn$ final state.
  In the non-mesonic $\QF$ process, a recoiled $\bar{K}$ is almost on-shell and absorbed by the two spectator nucleons.
  In $\QF$, $\q$ is predominantly defined by the neutron emission angle, because the residual nucleons are spectators (almost at-rest).
  Thus, the $\m$ distribution-centroid is given as 
  \begin{equation}
    \label{eq:quasifree-mass}
    M_{F}(\q) = \sqrt{ 4 m_{N}^{2} + m_{\bar{K}}^{2} + 4 m_{N} \sqrt{ m_{\bar{K}}^{2} + \q^{2} } },
  \end{equation}
  where $m_{N}$ and $m_{\bar{K}}$ are the intrinsic mass of $N$ and $\bar{K}$, respectively.
  We plotted the $M_{F}(\q)$-curve in Fig.~\ref{fig:im_vs_q} as a blue dotted line.
  In the figure, two event concentrations on $M_{F}(\q)$ are clearly seen around $\q \sim 0.2\ {\rm GeV}/c$ and $\sim 1.0\ {\rm GeV}/c$.
  These event concentrations correspond to the backward and forward scattered $\bar{K}$ in the elementary $K^{-}N \to \bar{K}n$ reaction.
  The $\QF$ should distribute around $M_{F}(\q)$ in the $\m$ direction due to the Fermi-motion of the two nucleons.
  To describe the distribution, a Gaussian function is utilized, as
  \begin{equation}
    \begin{split}
      \label{eq:quasi-free}
      f_{F} \left( \m, \q \right) &= \exp\!\! \left[ -\frac{ \left(\m-M_{F}(\q) \right)^{2} }{ \sigma^{2}(\q) } \right] \\
      &\quad \times \left[ A_{0}^{F}\,\exp\! \left( - \frac{\q^{2}}{Q_{F}^{2}} \right) + A_{1}^{F} \right. \\
      &\quad\quad\quad \left. + A_{2}^{F}\,\exp\! \left( \frac{\m}{m_{0}}+\frac{\q}{q_{0}} \right) \right].
    \end{split}
  \end{equation}
  In the formula, we allowed the $\m$ distribution width to have a $\q$ dependence as
  \begin{equation}
    \sigma(\q) = \sigma_{0} + \sigma_{2} \q^{2}.
  \end{equation}
  The second angle bracket in Eq.~\ref{eq:quasi-free} represents the $\q$ dependence of the production yield of the $\QF$ process, while the middle term is for flat distribution, and the first and third terms correspond to backward and forward scattered $\bar{K}$ events, respectively.

  The forward $\bar{K}$ part of the $\QF$ process is located far from the region of interest (distributed around the projectile $K^{-}$ momentum $\sim$ 1 GeV/$c$), as shown in Fig.~\ref{fig:im_vs_q} and Fig.~\ref{fig:fit_functions}-(b), so we phenomenologically formulated our model fitting function as an exponential for simplicity, as given in Eq.~\ref{eq:quasi-free}.

  \subsubsection{\label{sec:analysis:spectral-fitting:broad-distribution}
  Broad distribution {\rm (}$i=B${\rm )}
  }
  The two reaction processes described above have specific regions where events concentrate.
  However, there is a broad distribution, which cannot be explained easily, over the entire kinematically allowed region in $(\m,\q)$.
  In contrast to other processes, $\Lambda$, $p$, and $n$ share the kinetic energy rather randomly, resulting in a relatively weak $\m$ and $\q$ dependence, similar to a point-like interaction whose cross-section should be proportional to $\rho(\m,\q)$, and thus $f_{i}(\m,\q) \sim {\rm constant}$.
  A natural interpretation of this component is the three-nucleon absorption ($3\NA$) reaction of an incident $K^{-}$.
  On the other hand, there is a weak but yet clear $\m$ and $\q$ dependence over the whole kinematical region.
  The event density at higher $\m$ and lower $\q$ is much weaker than that at the opposite side.
  On the other hand, there is no clear event density correlation between $\m$ and $\q$, which indicates that the distribution could be described by the Cartesian product of centroid concentrating functions in both $\m$ and $\q$.
  The most natural formula can be written as an extension of Eq.~\ref{eq:bound-state} as
  \begin{equation}
    \begin{split}
      \label{eq:broad}
      f_{B} \left( \m,\q \right) &= \frac{(\Gamma_{B}/2)^{2}}{ \left(\m - M_{B} \right) ^{2} + \left( \Gamma_{B}/2 \right)^{2} } \\
      &\quad \times \left( A_{0}^{B} + A_{2}^{B} \frac{\q^{2}}{Q_{B}^{2}} \right) \,\exp\! \left( - \frac{\q^{2}}{Q_{B}^{2}} \right).
    \end{split}
  \end{equation}

  \subsubsection{\label{sec:analysis:spectral-fitting:m-spectra-of-Lpn-final-state}
    $\m$ spectra of $\Lambda pn$ final state
  }
\begin{figure}[h]
    \includegraphics[width=1.0\linewidth]{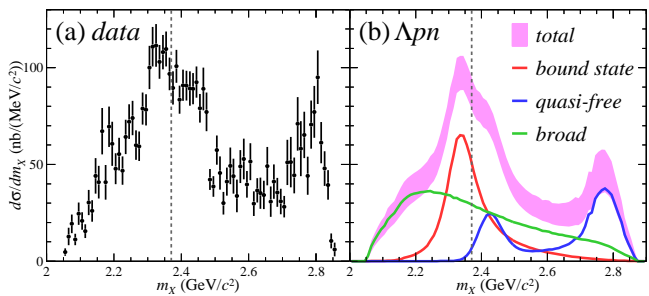}
    \caption{
      (color online) 
      $\m$ distribution (integrated by $\q$ over the whole kinematically allowed region) of (a) data and (b) model functions.
      The model function is limited to the $\Lambda pn$ final state; {\it i.e.}, the $\SNN$ contribution is excluded.
      The colored lines are spectra of three processes.
      The magenta thick curve is the sum of all the processes with an error band of the 95\% confidence level.
      The vertical gray dotted lines are $\MKNN$.
    }
    \label{fig:shape_lpn}
  \end{figure}

  To demonstrate the applicability of the model fitting functions conceptually, we present the $\m$ spectrum of the data in the $\Lambda pn$-selection window and compare it with the $\m$ spectral shapes, restricting ourselves to the $\Lambda pn$ final state, for $K$) $\KNN$, $F$) $\QF$, and $B$) the broad distribution, as shown in Fig.~\ref{fig:shape_lpn}.
  For comparison, the acceptance was corrected for the data Fig.~\ref{fig:shape_lpn}-(a) by dividing the data by $\varepsilon(\m,\q)$ bin by bin (except for $\varepsilon(\m,\q)<0.5\%$).
  For the same reason, weighting of the phase-space volume was applied to Fig.~\ref{fig:shape_lpn}-(b) by multiplying each function by $\rho(\m,\q)$.
  Both figures were integrated over the whole $\q$ region.
  All the parameters of the fitting functions of Fig.~\ref{fig:shape_lpn}-(b) were fixed to the final fitting value.
  For the figure, the 2D experimental resolution (depending on both $\m$ and $\q$) was considered in the Monte Carlo simulation.
  The magenta band is the sum of all the reaction components and the band width indicates the fit error.

  As shown in the figure, the global structure of the $\m$ spectrum is qualitatively described only with the $\Lambda pn$ final state, even before considering the $\SNN$ contribution, as expected.
  The quantitative fitting was performed by considering $\SNN$ effects, as described in the following section.

  \subsection{\label{sec:analysis:effect-of-SNN-contamination}
  Effect of $\bm{\SNN}$ contamination
  }
  \begin{figure*}[t]
    \includegraphics[width=0.7\linewidth]{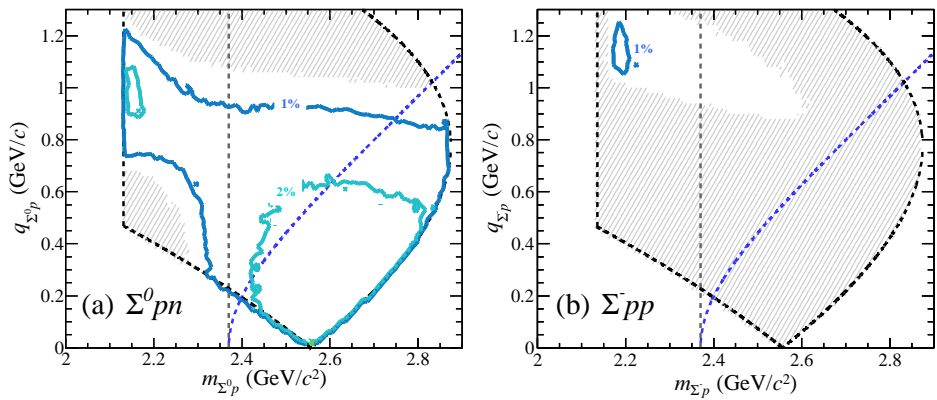}
    \caption{
      (color online) Experimental acceptance for each $\SNN$ contamination: (a) $\Sigma^{0} pn$ final state and (b) $\Sigma^{-}pp$ final state.
      The vertical and horizontal axes for $\Sigma^{0}pn$ ($\Sigma^{-}pp$) are the momentum transfer and invariant mass of the $\Sigma^{0}p$ ($\Sigma^{-}p$) system.
      The hatched regions are insensitive in the present setup, where $\varepsilon < 0.5\%$.
      The roughness of the contours is due to the limited statistics of the simulation. 
      The efficiency is calculated bin by bin.
      The vertical gray dotted lines and blue dotted curves are the same as in Fig.~\ref{fig:im_vs_q}.
    }
    \label{fig:im_vs_q_acce}
  \end{figure*}
  \begin{figure}[t]
    \includegraphics[width=1.0\linewidth]{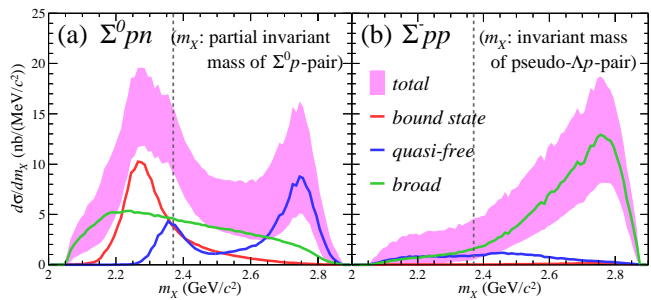}
    \caption{
      (color online) Expected spectral shapes of $\m$ for the (a) $\Sigma^{0}pn$ final state and (b) $\Sigma^{-}pp$ final state.
      The horizontal axis of (a) is the $\Lambda p$ invariant mass, which is a partial invariant mass of $\Sigma^{0}p$ where $\gamma$ of $\Sigma^{0} \to \gamma \Lambda$ is missing.
      The axis of (b) is the $\Lambda p$ invariant mass after miss-identification as the $\Lambda pn$ final state; {\it i.e.}, $X$ is a pair of pseudo-$\Lambda$ and $p$.
      The vertical gray dotted lines are $\MKNN$.
      Note that the full scale of the differential cross-section is different from that of Fig.~\ref{fig:shape_lpn}.
    }
    \label{fig:shape_snn}
  \end{figure}
  As we described in Sec.~\ref{sec:analysis:event-selection}, the selected $\Lambda pn$ events are not free from contamination from the $\SNN$ ($\Sigma^{0}pn$ and $\Sigma^{-}pp$) final states.
  The effect of these contaminations should be taken into account in generating the final spectral fitting.
  It is clear that an ideal method to evaluate the contaminations is to observe the $\SNN$ final state separately.
  Unfortunately, this is not possible with the present experimental setup.
  In the present analysis, we assumed the $\SNN$ channels are produced in analogue reaction processes with that of $\Lambda pn$, i.e., $K$) $\KNN$, $F$) $\QF$, and $B$) the broad distribution, and thus the same functions $f_{i}$ as the $\Lambda pn$ final state can be applied to represent the $(\m,\q)$ event distribution of the $\SNN$ final states.
  When $f_{i}$ and its parameters are given as a common function, the $\YNN$ final states and their contributions to the spectra through the $\Lambda pn$-selection window can be reliably evaluated by expanding $F_{i}$ to $F_{i}^{j}$ so the formula is also applicable to $\SNN$, where $j = (\Lambda pn, \Sigma^{0}pn, \Sigma^{-}pp)$ and $F_{i}^{j} = \rho_{j}\, f_{i}$.
  $\rho$ and $\varepsilon$ can also be expanded to account for each final state in the same manner.

  For the $\Sigma^{0}pn$ final state, $X$ is produced in the same way as the $\Lambda pn$ final state, but $X$ goes to $\Sigma^{0} p$ instead of $\Lambda p$.
  Because $\Sigma^{0}$ decays to $\gamma \Lambda$ (100\%), part of the $\Sigma^{0} pn$ final state leaks in the $\Lambda pn$-selection window.
  As shown in Fig.~\ref{fig:im_vs_q_acce}-(a), the simulated acceptance over $(m_{\Sigma^{0}p}, q_{\Sigma^{0}p})$ is smaller but similar to Fig.~\ref{fig:im_vs_q_effic}-(b).
  The expected $\m$ and $\q$ for the contaminating events are also simulated, and the resulting $\m$ spectrum is shown in Fig.~\ref{fig:shape_snn}-(a).
  As shown in the figure, the structure in the spectrum is similar but shifted to the lower side compared to Fig.~\ref{fig:shape_lpn}-(b), due to the missing energy of the $\gamma$-ray.

  In contrast, the situation is very different for the $\Sigma^{-}pp$ final state.
  We simulated this channel in a similar manner to that used for the $\Sigma^{0} pn$ final state by replacing a $\Sigma^{0}p$-pair with a $\Sigma^{-}p$-pair.
  The $\Sigma^{-}$ decays to $n\pi^{-}$ ($\sim$ 100\%).
  When the invariant mass of the $\pi^{-}$ and one of the protons in this final state happen to be close to the $\Lambda$ intrinsic mass, the event may enter the $\Lambda pn$-selection window.
  This makes the simulated acceptance over $(m_{\Sigma^{-}p}, q_{\Sigma^{-}p})$ given in Fig.~\ref{fig:im_vs_q_acce}-(b) very different from the other two.

  We simulated $\m$ and $\q$ of the contaminated events for the incorrect $\Lambda p$-pair (pseudo-$\Lambda p$-pair), which would be analyzed as the $\Lambda pn$ final state in the analysis code.
  The resulting $\m$ spectrum is given in Fig.~\ref{fig:shape_snn}-(b).
  As shown in the figure, the structure in the spectrum is also totally different from the other $\m$ spectra.
  It should be noted that we generated the $I_{z}=-1/2$ $\KNN$ ($\bar{K}^{0}nn$) bound state instead of $I_{z}=+1/2$ in this $\Sigma^{-}pp$ simulation at the same relative yield with the other two final states.
  This assumption might not be valid, because the isospin combination in the formation channel is different.
  However, it does not affect the fitting, because events from $\KNN$ concentrate at the lower $\q$-side, as shown in Fig.~\ref{fig:fit_functions}-(a), where our detector system does not have sensitivity for the $\Sigma^{-}pp$ final state, as shown by the hatched region in Fig.~\ref{fig:im_vs_q_acce}-(b).
  For the same reason, the contribution from the $\QF$ process to this final state is much smaller than those in the other final states.
  \subsection{\label{sec:analysis:iterative-fitting-procedure}
  Iterative fitting procedure
  }
  To determine the spectroscopic parameters, we conducted 2D fitting for the 2D event distribution, as described in Ref.~\cite{ajimura.plb.2019}.
  As shown in Fig.~\ref{fig:im_vs_q_effic}-(a) by the gray hatching, the present setup has insensitive regions due to the geometrical coverage of the CDS.
  To avoid spurious bias caused by the acceptance correction, we directly compared the data and the fitting function in the count base
  by computing the expected event-numbers $\lambda(\m,\q)$ to be observed in a $(\m,\q)$-bin by
  \begin{equation}
    \begin{split}
      \lambda(\m,\q) = \sum_{i,j} R_{j}\, \varepsilon_{j}(\m,\q)\, F_{i}^{j}(\m,\q)\, \Delta \m\, \Delta \q \\
      = \sum_{i,j} R_{j}\, \varepsilon_{j}(\m,\q)\, \rho_{j}(\m,\q)\, f_{i}(\m,\q)\, \Delta \m\, \Delta \q, 
    \end{split}
  \end{equation}
  where $\Delta \m$ and $\Delta \q$ are the bin widths.
  Then, we evaluated the probability of observing data in the $(\m,\q)$-bin as $P(Z=N(\m,\q))$, where $P$ is the Poisson distribution function, $N(\m,\q)$ is the data counts at the $(\m,\q)$-bin, and $Z$ is a random Poisson variable for the expectation value of $\lambda(\m,\q)$.
  The log-likelihood for the 2D fitting $ln.L$ can be defined as an ensemble of probabilities as
  \begin{equation}
    \label{eq:fit_lnl}
    ln.L = -\sum_{\m,\q} \ln(P(Z=N(\m,\q))),
  \end{equation}
  and the maximum $ln.L$ was obtained to fit the data by optimizing the spectroscopic parameters.
  There are a total of 17 parameters in this fitting, consisting of four parameters for the $\KNN$ bound state, eight parameters for the non-mesonic $\QF$ process, and five parameters for the broad component.
  For the summation for $ln.L$, we omitted the $(\m,\q)$-bin having no statistical significance where $\varepsilon_{j}(\m,\q)<0.5\%$.

  It is very important to apply the acceptance correction to properly represent the physics behind the system.
  It is also true that the spectra cannot be presented in the scale of the cross-section.
  Therefore, we applied acceptance correction for the events in the $\Lambda pn$-selection window after the fitting procedure converged by dividing the spectra by $\varepsilon_{\Lambda pn}(\m,\q)$ bin by bin for both the data and fit results, except for Figs.~\ref{fig:mm_vs_lnl}-\ref{fig:mmlnl}.

  Due to the asymmetrical kinematical limits (see Fig.~\ref{fig:im_vs_q}), the spectral function largely depends on the $\q$-region.
  We performed a first fitting for the whole region as the global fit, then performed a second fitting for only the $\q$ region from 0.3 to 0.6 GeV/$c$ to focus on $\KNN$.
  The second fitting was conducted to deduce the parameters of $\KNN$ under a better S/N region, so the other parameters are fixed in the second fitting.
  After an iteration of a spectral fitting for the data shown in Fig.~\ref{fig:im_vs_q}, we looped back to evaluate the ratio of the final state yields of $\Lambda pn : \Sigma^{0}pn : \Sigma^{-}pp : \rm{\it{other}}$ in the $\Lambda pn$-selection window by the fitting procedure described in Sec.~\ref{sec:analysis:event-selection} (see Fig.~\ref{fig:mmlnl} and Tab.~\ref{tab:relative-yield}).
  To obtain self-consistent results, we looped back over the two procedures iteratively until both the ratio parameters and spectroscopic parameters converge.

  \section{\label{sec:results-and-discussion}
  Results and Discussion
  }

  \subsection{\label{sec:results-and-discussion:2d-fitted-spectra}
  2D fitted spectra
  }
\begin{figure}[t]
    \includegraphics[width=1.0\linewidth]{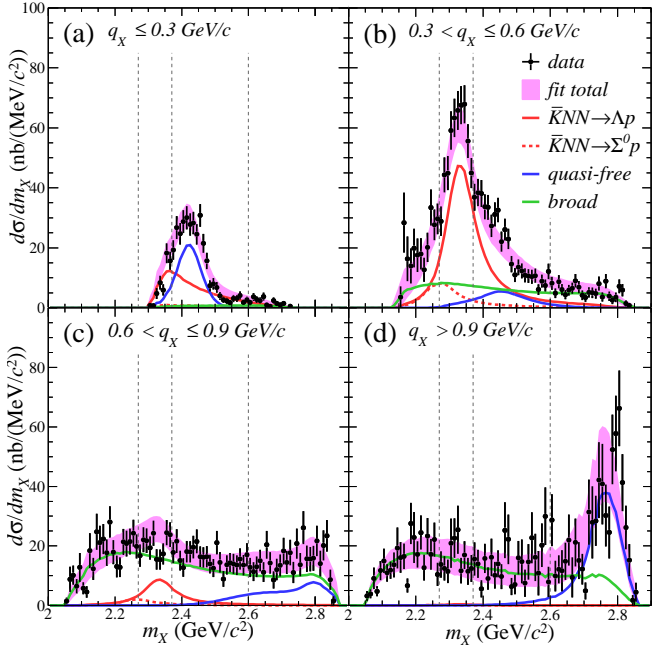}
    \caption{
      (color online) $\m$ spectra for various intervals of $\q$: (a) $\q \leq 0.3\ {\rm GeV}/c$, (b) $0.3< \q \leq 0.6\ {\rm GeV}/c$, (c) $0.6< \q \leq 0.9\ {\rm GeV}/c$, and (d) $0.9\ {\rm GeV}/c < \q$.
      The dotted lines correspond to the $\m$-slice regions given in Fig.~\ref{fig:cs_momtrans}.
    }
    \label{fig:cs_mass}

  \end{figure}

\begin{figure}[t]
    \includegraphics[width=1.0\linewidth]{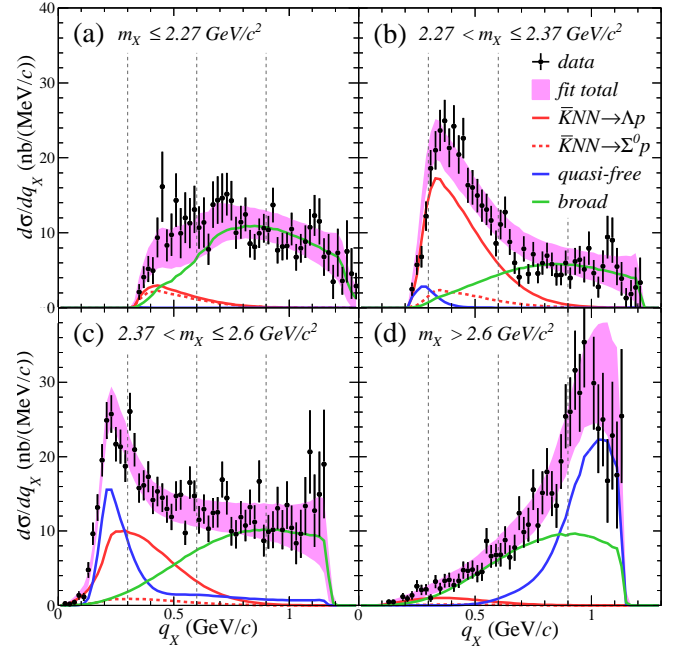}
    \caption{
      (color online) $\q$ spectra for various intervals of $\m$: (a) $\m \leq 2.27\ {\rm GeV}/c^{2}$, (b) $2.27< \m \leq 2.37\ {\rm GeV}/c^{2}$, (c) $2.37< \m \leq 2.6\ {\rm GeV}/c^{2}$, and (d) $2.6\ {\rm GeV}/c^{2}<\m$, with the fitting results shown as colored lines.
      The dotted lines correspond to the $\q$-slice regions given in Fig.~\ref{fig:cs_mass}.
    }
    \label{fig:cs_momtrans}
  \end{figure}

  To demonstrate the accuracy of the fit result in 2D, we plotted the fit result for the $\m$-spectra in the $\q$-slice (as shown in Fig.~\ref{fig:cs_mass}) and for the $\q$-spectra in the $\m$-slice (as shown in Fig.~\ref{fig:cs_momtrans}), $i.e.,$ projections of 2D data onto the $\m$-axis and $\q$-axis at the same time.
  In other words, Figs.~\ref{fig:cs_mass} and \ref{fig:cs_momtrans} show the compilation of event projections of the two-dimensional four-by-four $\m$- and $\q$-regions of Fig.~\ref{fig:im_vs_q} onto each axis.
  In each spectrum, data are compared with the fit result as shown in the magenta band (95\% confidence level), and decomposed as colored lines.
  All the regions are well reproduced for both the $\m$ and $\q$ spectra. 
  The maximum log-likelihood and total number of degrees of freedom of the fitting were 2425 and 2234, respectively.
  We plotted the signal of $\KNN$ formation and its $\Lambda p$ decay as a red line, and $\KNN \to \Sigma^{0}p$ in the $\Lambda pn$-selection window as a red dashed-line.
  To simplify the plot, we summed the $\QF$ and broad contributions from the $\Lambda pn$ final state and from contaminations of the $\SNN$ final states, because the spectra for each reaction process are relatively similar (see Figs.~\ref{fig:shape_lpn}-(b) and \ref{fig:shape_snn}).
  As expected, the $\KNN$ formation signal is clearly seen in Fig.~\ref{fig:cs_mass}-(b) in the $\m$ spectrum, and in Fig.~\ref{fig:cs_momtrans}-(b) in the $\q$ spectrum.

  At the lowest $\q$ region of the $\m$ spectrum in Fig.~\ref{fig:cs_mass}-(a), the spectrum is confined in a medium mass region due to the kinematical boundary (see Fig.~\ref{fig:im_vs_q} and Fig.~\ref{fig:im_vs_q_effic}).
  In this region, the backward $\bar{K}$ part of the $\QF$ process $K^{-}+N \to \bar{K}+n$ becomes dominant.
  In  Fig.~\ref{fig:cs_mass}-(b), the $\KNN$ formation signal is dominant and contributions from other processes, in particular the $\QF$ process, are relatively suppressed.
  In the relatively large $\q$ region in Fig.~\ref{fig:cs_mass}-(c), the broad component becomes dominant, while the $\KNN$ formation signal becomes weaker.
  At an even larger $\q$ region in Fig.~\ref{fig:cs_mass}-(d), the forward $\bar{K}$ part of the $\QF$ process becomes large, which distributes to the large $\m$ side.
  This events concentration may partially arise from direct $K^{-}$ absorption on two protons in $^{3}{\rm He}$ ($2\NA$), but the width is too great to be explained by the Fermi motion.
  Therefore, it is difficult to interpret $2\NA$ as the dominant process of this events concentration.
  In this $\q$ region, there is also a large contribution from the broad component.

  Figure~\ref{fig:cs_momtrans} shows the $\q$ spectra sliced on $\m$.
  Figure~\ref{fig:cs_momtrans}-(a) shows the region below the $\KNN$ formation signal where the broad distribution is dominant, having small leakage from the signal.
  As shown in the spectrum, the broad distribution has no clear structure and has a larger yield at a higher $\q$ region than at a lower $\q$ region.
  Figure~\ref{fig:cs_momtrans}-(b) shows the $\KNN$ formation signal region, in which the events clearly concentrate at the lower $\q$ side.
  In Fig.~\ref{fig:cs_momtrans}-(c), we can see the backward $\bar{K}$ part of the $\QF$ process, together with the leakage from the signal and broad distribution.
  In contrast to $\KNN$, the $\QF$ process even more strongly concentrates in the lower $\q$ region (neutron is emitted to the very forward direction).
  To compare the $\q$ dependence with that of the $\KNN$ formation process, we formulated our model fitting function for the forward $\bar{K}$ $\QF$ process to have a Gaussian form (see Eq.~\ref{eq:quasi-free}).
  The $\q$ spectrum at the highest $\m$ region is given in Fig.~\ref{fig:cs_momtrans}-(d).
  The major components are the broad distribution and the forward $\bar{K}$ part of the $\QF$ process.
  The centroid of the event concentration locates at an incident kaon momentum of 1 GeV/$c$, but the width in $q$ is again too great to interpret it as being due to the $2\NA$ reaction.
  Thus, the $2\NA$ process would be rather small in the case of the $\Lambda pn$ final state of the present reaction.

\begin{figure}[h]
    \includegraphics[width=1.0\linewidth]{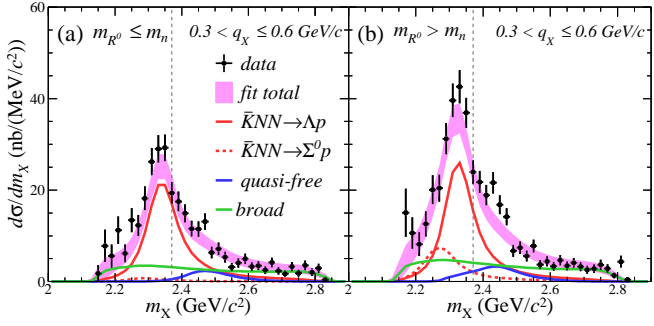}
    \caption{
      (color online) $\m$ spectra for (a) $m_{R^{0}} \leq m_{n}$ and (b) $m_{R^{0}} > m_{n}$, with the fitting result.
      The selected $q_{X}$ region is the same as for Fig.~\ref{fig:cs_mass}-(b) ($0.3 < \q \leq 0.6\ {\rm GeV}/c$).
      The vertical dotted line is $\MKNN$.
      Note that the bin width is exceptional, $\Delta \m = 20$ MeV/$c^{2}$, for these figures to have sufficient statistics.    }
    \label{fig:mass_mmcut}
  \end{figure}

  To check the $\Sigma^{0}pn$ contamination effect in the present fitting, we divided Fig.~\ref{fig:cs_mass}-(b) into two regions for $m_{R^{0}} \leq m_{n}$ and $m_{R^{0}} > m_{n}$, as shown in Fig.~\ref{fig:mass_mmcut}.
  The figure shows that the spectra are consistent with the $\Sigma^{0}pn$ final state distribution in Fig.~\ref{fig:mmlnl}-(a), $i.e.,$ that the $\KNN \to \Sigma^{0}p$ contribution exists only on the $m_{R^{0}} > m_{n}$ side.
  As shown in the figure, the $\m$ spectrum of Fig.~\ref{fig:mass_mmcut}-(b) below the mass threshold of $\MKNN$ is slightly wider and deeper than that of Fig.~\ref{fig:mass_mmcut}-(a) in both the data and total fitting function, as expected, due to the presence of $\Sigma^{0}pn$ contamination.

  \subsection{\label{sec:results-and-discussion}
  Fitted parameters
  }
  The converged 17 spectroscopic parameters are listed in Tab.~\ref{tab:fit-result}.
  We improved the fitting procedure to fully take into account the $\SNN$ final states in the present analysis, as well as the $(\m,\q)$ dependence of the detector resolution.
  As a result, the values of the spectroscopic parameters were updated from our recent publication \cite{ajimura.plb.2019},
  though the updated values are within the error range of the previous publication.

  \begin{table}[h]
    \caption{
      Converged 17 spectroscopic parameters and their errors. }
    \label{tab:fit-result}
    \renewcommand{\arraystretch}{1.3}
    \begin{tabularx}{\linewidth}{CD{,}{\pm}{18}} \hline \hline
      $\KNN$ bound state	  	&	{\rm Value},({\rm stat.})^{+}_{-}({\rm syst.})					\\
      \hline
      $A_{0}^{K}$							&(1.523,0.103^{+0.001}_{-0.119})\times10^{4}				\\
      $M_{K}$									&	2.328,0.003^{+0.004}_{-0.003}\ {\rm GeV}/c^{2}		\\
      $\Gamma_{K}$						&	0.100,0.007^{+0.019}_{-0.009}\ {\rm GeV}					\\
      $Q_{K}$									&	0.383,0.011^{+0.004}_{-0.001}\ {\rm GeV}/c				\\
      \hline
      \hline
      Non-mesonic $\QF$       & {\rm Value},({\rm stat.})^{+}_{-}({\rm syst.})					\\
      \hline
      $A_{0}^{F}$							&(4.045,0.800^{+0.408}_{-0.953})\times10^{4}				\\
      $A_{1}^{F}$							&(1.496,0.416^{+0.662}_{-0.456})\times10^{2}				\\
      $A_{2}^{F}$							&(2.947,0.292^{+0.000}_{-2.947})\times10^{-39}			\\
      $\sigma_{0}$						&	0.045,0.004^{+0.006}_{-0.001}\ {\rm GeV}/c^{2}		\\
      $\sigma_{2}$						&	0.169,0.053^{+0.047}_{-0.037}\ {\rm GeV^{-1}}			\\
      $Q_{F}$									&	0.172,0.009^{+0.003}_{-0.005}\ {\rm GeV}/c				\\
      $m_{0}$									&	33.66,0.047^{+0.301}_{-1.641}\ {\rm MeV}/c^{2}		\\
      $q_{0}$									&	72.81,0.644^{+0.162}_{-9.590}\ {\rm MeV}/c				\\
      \hline
      \hline
      Broad distribution      &	{\rm Value},({\rm stat.})^{+}_{-}({\rm syst.})					\\
      \hline
      $A_{0}^{B}$							&(0.596,2338^{+3767}_{-0.000})\times10^{-12}			\\
      $A_{2}^{B}$							&(2.924,0.408^{+0.000}_{-0.104})\times10^{3} 				\\
      $M_{B}$									&	2.128,0.032^{+0.013}_{-0.000}\ {\rm GeV}/c^{2}		\\
      $\Gamma_{B}$						&	0.532,0.068^{+0.000}_{-0.031}\ {\rm GeV}					\\
      $Q_{B}$									&	0.689,0.066^{+0.011}_{-0.001}\ {\rm GeV}/c				\\
      \hline \hline
    \end{tabularx}
    \renewcommand{\arraystretch}{1}
  \end{table}

  The mass position of the $\KNN$ bound state $M_{K}$ (or the binding energy $B_{K}\equiv \MKNN-M_{K}$) and its decay width $\Gamma_{K}$ are
  \begin{equation*}
    \begin{split}
      M_{K} = \valM \\
      (B_{K} = \valB),
    \end{split}
  \end{equation*}
  \begin{equation*}
    \Gamma_{K} = \valG,
  \end{equation*}
  respectively.
  The $S$-wave Gaussian reaction form factor parameter of the $\KNN$ bound state $Q_{K}$ is
  \begin{equation*}
    Q_{K} = \valQ.
  \end{equation*}
  The total production cross-section of the $\KNN$ bound state going to the $\Lambda p$ decay mode $\sigma^{tot}_{K} \cdot \BR_{\Lambda p} $ was evaluated by integrating the spectrum to be
  \begin{equation*}
    \sigma^{tot}_{K} \cdot \BR_{\Lambda p} = \valCSLp.
  \end{equation*}

  In the present analysis, the strength of the $\KNN \to \Sigma^{0}p$ decay mode is deduced based on the $\Sigma^{0}pn$ contamination yield given by Fig.~\ref{fig:mmlnl}.
  By assuming that the relative yields of the three physical processes of $\SNN$ and those of $\Lambda pn$ are equal, we estimated the differential cross-section of $\KNN$ decaying into the $\Sigma^{0}p$ mode $\sigma^{tot}_{K} \cdot \BR_{\Sigma^{0} p}$ as
  \begin{equation*}
    \sigma^{tot}_{K} \cdot \BR_{\Sigma^{0} p} = \valCSSp.
  \end{equation*}
  Therefore, the branching ratio of the $\Lambda p$ and $\Sigma^{0}p$ decay modes was estimated to be $ \BR_{\Lambda p}/\BR_{\Sigma^{0}p} \sim \valBR$.
  The estimated branching ratio is higher than the value of the theoretical calculation based on the chiral unitary approach, predicting a ratio of almost one \cite{sekihara.prc.86}.

  \subsection{\label{sec:results-and-discussion:systematic-errors}
  Systematic errors
  }
  The systematic errors were evaluated by considering the uncertainties of the absolute magnetic field strength of the solenoid, the binning effect of spectra, and systematic errors of the branch of the final states (Tab.~\ref{tab:relative-yield}).
  For production cross-sections, we considered the luminosity uncertainty.
  To be conservative, the evaluated systematic errors are added linearly.

  We succeeded in reproducing the data distribution by our model fitting functions.
  However, for the broad distribution, we cannot simply specify the physical process of its formation.
  Thus, we also tried an independent model fitting functions, which are intentionally unphysical but still able to reproduce the global data structure.
  A typical model fitting function fulfilling the requirements can be obtained by replacing the $\q$-even polynomial term with a simple $\q$-proportional one in Eq.~\ref{eq:broad}.
  The $\q$-proportional term is not physical by itself, and can only be possible as a comprehensive interference of an $S$-wave and a $P$-wave.
  As yet another extreme of the model fitting function of the broad distribution, we also examined a fit by replacing Loretnzian term of Eq.~\ref{eq:broad} to the second order polynomials.
  Although these alternative model functions are unphysical, we treated the centroid shifts of the other parameters as a source of systematic error for safety.

  The systematic uncertainties are much reduced from Ref.~\cite{ajimura.plb.2019},
  due to the improved analysis procedure by considering a precise and realistic evaluation of the $\SNN$ contamination in the $\Lambda pn$-selection window.

  \subsection{\label{sec:results-and-discussion:discussion}
  Discussion
  }
  We introduced three physical processes to account for the data, $K$) $\KNN$ state production, $F$) $\QF$ process, and $B$) the broad distribution, and found that the presence of $K$) $\KNN$ is essential to explain the spectra self-consistently, which cannot be formed as an artifact.
  The presence of $F$) is naturally expected from the analysis on inclusive channel presented in Ref.~\cite{hashimoto.ptep.2015}, but the relative yield of the quasi-free component is substantially reduced because we focused on the non-mesonic $\Lambda pn$ final state in the present paper.
  For process $B$), we pointed out that the possibility that it could be due to point-like $3\NA$ kaon absorption, because of the weakness of its $(\m, \q)$ dependence.

  For the $\KNN$ bound state, $B_{K}\sim40\ {\rm MeV}$  agrees nicely with phenomenological predictions\cite{yamazaki.prc.2007,ikeda.prc.2007,ikeda.prc.2009,revai.prc.2014,wycech.prc.2009}.
  However, it should be noted that the obtained $B_{K}$ is the spectral Breit--Wigner pole position, neglecting the microscopic reaction dynamics given in Eq.~\ref{eq:X-formation}. 
  Thus, the present Breit--Wigner pole might be different from the physical pole predicted by theoretical calculations.

  $\Gamma_{K}\sim100\ {\rm MeV}$ is wide, as for a quasi-bound state, compared to the binding energy $B_{K}$.
  It is also wider than the $\Lambda (1405) \to \pi \Sigma$ decay width of $\sim$ 50 MeV (100\%).
  If $\Lambda (1405)$ is the $\KN$ quasi-bound state, then it is naturally expected that the $\KNN \to \pi \Sigma N$ decay will occur in the same order as the $\YN$ decay channels.

  As shown in Fig.~\ref{fig:cs_momtrans}-(b), the production yield of the $\KNN$ bound state is much larger in a smaller $\q$ region.
  This trend is a common feature of nuclear bound-state formation reactions in general.
  In the $K^{-} + ~^{3}{\rm He} \to X + n$ formation channel, we can achieve a minimum momentum transfer to $X$ as small as $\sim$ 200 MeV/$c$, which makes this channel the ideal formation process.
  However, $\sigma^{tot}_{K} \cdot \BR_{\YN~(=\Lambda p,\, {\rm or}\, \Sigma^{0}\!p)}$ is still small compared to the total cross-section of the elementary $K^{-}N \to \bar{K}n$ reaction by the order of $\mathcal{O}(10^{-3})$.
  Even if we take into account a mesonic decay branch similar to $\YN$ decay, the total $\KNN$ formation branch would still be less than $\mathcal{O}(10^{-2})$ of the elementary cross-section.
  In spite of the small formation yield and large decay width near the binding threshold, we have succeeded in observing kaonic bound state formation.
  This is because the $\YNN$ final states, which strongly limit the number of possible complicated intermediate states such as mesonic processes, allow the $s$-quark flow in the reaction to be traced by $Y$, and moreover, the $\KNN$ signal and remaining non-mesonic $\QF$ processes can be effectively separated by $\q$-slicing.

  Let us consider the physical meaning of $Q_{K}$ in Eq.~\ref{eq:bound-state}.
  $Q_{K}$ is quite large, more than twice the $Q_{F}$ of the non-mesonic $\QF$ process.
  The value of $Q_{F}$ is natural in view of the size of the $^{3}{\rm He}$ radius, as well as the strong angular dependence of the elementary process $K^{-}N \to \bar{K}n$ observed in Ref.~\cite{hashimoto.ptep.2015} at $p_{K^{-}} =$ 1 GeV/$c$, which is the primary reaction of Eq.~\ref{eq:X-formation}.
  Instead, the value of $Q_{K}$ may carry information on the spatial size of the $\KNN$ state.
  We formulated the model fitting function based on a simple PWIA calculation, assuming that the $\KNN$ wave function can be written in the ground state of a harmonic oscillator (HO).
  The spatial size of the HO wave function can be given as $R_{K} = \hbar / Q_{K} \sim 0.5$ fm (if we take into account the correction factor of the c.m.~motion, $(2 m_{N} + m_{\bar{K}})/ 2m_{N}$, $R_{K} \sim 0.6$ fm).
  The compactness is also naively supported by the large $B_{K} \sim$ 40 MeV.

  Finally, we briefly discuss the broad component.
  The present data show that the $2\NA$ kaon absorption channels are weak, in contrast to kaon absorption at-rest experiments \cite{delgrande.epjc.79}, so we need to understand why $3\NA$ still exists while the $2\NA$ channels are weak.
  The distribution of this component $f_{B}$, given in Fig.~\ref{fig:fit_functions}-(c), becomes a broad $P$-wave resonance-like structure characterized by $M_{B}$ between $m_{\Lambda} + m_{p}$ and $\MKNN$, $A_{0}^{B} << A_{1}^{B}$, as shown in Tab.~\ref{tab:fit-result}.
  This phenomenon might be simply due to the nature of the formula of the fitting function, given in Eq.~\ref{eq:broad}, but it is worth studying in more detail to clarify the physics of  this component.
  To be conservative, we keep our interpretation open for the physical process of this broad distribution, and treated that as a source of the systematic error.

  Open questions still remain, such as the spin-parity $J^{P}$ of the $\KNN$ state, and the relationship between the present $\KNN$ signal and $\Lambda(1405)$ resonance.
  Also, in the analysis, we have not taken into account the interference effects between the three introduced physical processes.
  More comprehensive studies are required to clarify these remaining questions.

  \section{\label{sec:summary}
  Summary
  }
  We have measured the $\Lambda pn$ final state in the in-flight reaction on a $^{3}{\rm He}$ target at a kaon momentum of $1\ {\rm GeV}/c$.
  We observed the kaonic nuclear quasi-bound state, $I_{z}=+1/2$ $\KNN$, and obtained its parameters by 2D fitting of the $\Lambda p$ invariant mass and momentum transfer.

  The binding energy and the decay width of the state were $B_{K} = \valB$ and $\Gamma_{K} = \valG$, respectively.
  The $S$-wave Gaussian reaction form-factor was $Q_{K} = \valQ$.
  The total production cross-sections of the $\KNN$ bound state decaying into non-mesonic $\Lambda p$ and $\Sigma^{0}p$ modes were obtained to be $\sigma^{tot}_{K} \cdot \BR_{\Lambda p} = \valCSLp$ and $\sigma^{tot}_{K} \cdot \BR_{\Sigma^{0} p} = \valCSSp$, respectively.
  Thus, the ratio $\Lambda p / \Sigma^{0}p$ decay branch was approximately $\valBR$.

  Although it would be premature to make a conclusion regarding the spatial size of $\KNN$ from a simple PWIA-based model fitting function, the implied size is quite small compared to the mean nucleon distance in normal nuclei.
  However, the observed value of $Q_{K} = \valQ$ is unexpectedly large (about twice as large as an elementary process), which makes the theoretical microscopic study difficult.
  Therefore, a more realistic theoretical calculation including detailed reaction dynamics and a more detailed experimental study are essential to understand the observed $\q$ distribution.

  \begin{acknowledgments}
    The authors are grateful to the staff members of J-PARC/KEK for their extensive efforts, especially on the stable operation of the facility.
    We are also grateful to the contributions of Professors Daisuke Jido, Takayasu Sekihara, Dr.~Rie Murayama, and Dr.~Ken Suzuki.
    This work is partly  supported by MEXT Grants-in-Aid 26800158, 17K05481, 26287057, 24105003, 14102005, 17070007, and 18H05402.
    Part of this work is supported by the Ministero degli Affari Esteri e della Cooperazione Internazionale, Direzione Generale per la Promorzione del Sistema Paese (MAECI), StrangeMatter project.
  \end{acknowledgments}

  \bibliography{bibfile}

\end{document}